\newcommand{\be}{\begin{eqnarray}}
\newcommand{\ee}{\end{eqnarray}}
\newcommand{\bee}{\begin{eqnarray*}}
\newcommand{\eee}{\end{eqnarray*}}

\def\r#1{\right#1}
\def\fr{\frac{1}{2}}

\def\mref#1{(\ref{#1})}
\def\rr{\mbox{\Bbb R}}
\def\p{\partial}

\def\bd{\begin{displaymath}}
\def\ed{\end{displaymath}}

\def\ba#1{\begin{array}{#1}}
\def\ea{\end{array}}
\def\nn{\nonumber}
\newfont{\Bbb}{msbm10 scaled 1200}
\renewcommand{\phi}{\varphi}
\documentclass[12pt]{article}
\usepackage{fleqn}
\title{On Calogero wave functions}
\author{C.Gonera\thanks{supported by the KBN grant 2 P03B 076 10}\ \\
P.Kosi\'nski$^*$\\
M.Majewski$^*$\\
P.Ma\'slanka$^*$\\
                              Theoretical Physics Department II\\
                                     University of Lodz\\
                          ul. Pomorska 149/153, 90 236 Lodz, Poland\\
}
\date{}
\begin{document}
\maketitle
\begin{abstract}
Two properties of Calogero wave functions for rational Calogero models
are
studied: (i)~the representation of the wave functions in terms of the 
exponential of Lassalle operators, (ii)~the $sL(2,\rr)$\ structure of
the Calogero--Moser wave functions.
\end{abstract}

\section{Introduction\label{s1}}
It is now well known that the Calogero model \cite{b1} is
superintegrable,
both in classical~\cite{b2} as well as in quantum version~\cite{b3}.
Superintegrability means here the existence of $2N-1$\ independent,
regular and globally defined integrals of motion which do not depend
explicitly on time ($N$\ is a number of degrees of freedom).

A superintegrable system has a number of interesting properties.
Its hamiltonian depends on the specific combinations of action 
variables~\cite{b4} which, on the quantum level, implies considerable
energy
degeneracy; angle--action variables are not defined uniquely. The latter
property has again its quantum counterpart: there is a freedom in the
choice of
basis diagonalizing the hamiltonian (this freedom is, of course, closely 
related to the energy degeneracy). Given a specific choice of
angle--action
variables one can take a basis spanned by common eigenvectors of the
quantum 
counterparts of action variables.

In the case of Calogero model few bases diagonalizing different sets of
commuting integrals of motion have been constructed. First, there exists
a basis spanned by the so--called Hi--Jack polynomials~\cite{b5};
the corresponding integrals are most simply expressible in the Dunkl
operator~\cite{b6} formalism. Second example of basis has been given by
Brink
et.al.\cite{b7}; again the relevant set of commuting integrals can be
easily
identified.

In both cases the basic wave functions can be related to known 
po\-ly\-no\-mials~\cite{b8}$\div$\cite{b11}. To this end one proceeds as
follows~\cite{b9}$\div$\cite{b13}. Consider the hamiltonian describing
the Calogero model
\be
H&=&\fr\sum\limits_i(p_i^2+\omega^2x^2_i)+{a(a-1)\over
1}\sum\limits_{i\not=j}
{1\over(x_i-x_j)^2}\label{w1}
\ee

The ground--state wave function and energy read, respectively  
\be
\Psi_0&=&\prod\limits_{i<j}(x_i-x_j)^a\exp\left(-{\omega\over
2}\sum\limits_i
x_i^2\r)\label{w2a}\\
E_0&=&{\omega N\over 2}\left((N-1)a+1\r);\label{w2b}
\ee
here $\Psi_0$\ is given in the sector $x_1\geq x_2\geq\ldots\geq
x_N$--the
extension to the other sectors depends on statistics. After gauging out
the
wave function $\Psi_0$\ one obtains
\be
\tilde{H}&=&\Psi_0^{-1}(H-E_0)\Psi_0=\omega\sum\limits_ix_i{\p\over\p
x_i}-
\fr\left(\sum\limits_i{\p^2\over \p x_i^2}+\right.\nn\\
&&\left.+a\sum\limits_{x\not= j}{1\over x_i-x_j}
({\p\over\p x_i}-{\p\over\p x_j})\r)
\equiv\omega\sum\limits_i x_i{\p\over\p x_i}-\fr O_L\label{w3}
\ee
$\tilde{H}$\ can be further reduced due to the identity
\be
e^{{1\over 4\omega}O_L}\tilde{H}e^{-{1\over 4\omega}O_L}&=&\omega
\sum\limits_ix_i{\p\over \p x_i}\label{w4}
\ee
Therefore, by a similarity transformation $\tilde{H}$\ reduces to the 
dilatation operator. This implies that any energy eigenfunction can
be written as a~product of ground--state wave function and a~polynomial
\be
\tilde{W}(x)&=&e^{-{1\over 4\omega}O_L}W(x)\label{w5}
\ee
where $W(x)$\ is a~symmetric homogeneous polynomial of degree $k$.

In order to obtain the basis spanned by the Hi--Jack polynomials one
takes $W(x)$\ to be Jack polynomial~\cite{b14}. On the other hand, Brink
et.al. basis corresponds to $W(x)$\ being monomial symmetric 
function~\cite{b13}

Historically, the first complete set of eigenfunctions was given by 
Ca\-lo\-ge\-ro~\cite{b1}. It can be written in the form (neglecting the 
ground--state factor)
\be
\phi_{nk}&=&L_n^b(\omega r^2)P_k(x)\label{w6}
\ee
where $L_n^b$\ is a~Laguerre polynomial, $P_k(x)$\ is translationally
invariant symmetric homogeneous polynomial of degree $k$\ (Calogero
polynomial) obeying
\be
O_LP_k(x)=0\label{w7}
\ee
and
\be
b&\equiv&k+\fr(N-3)+{N\over 2}(N-1)a\label{w8}\\
r^2&\equiv&{1\over 2N}\sum\limits_{i\not=j}(x_i-x_j)^2\label{w9}
\ee

The structure of wave functions, given by eqs.\mref{w6},\mref{w7}
results from
$sL(2,\rr)$\ dynamical symmetry inherent in the model~\cite{b15}. The
tower of
states obtained by the fixing $k$\ and varying $n$\ spans an irreducible
representation of $sL(2,\rr)$, eq.\mref{w7} being the condition for 
lowest--weight vector.

In the present paper we analyse some formal properties of Calogero
basis.
First, we construct the representation~\mref{w5} for Calogero wave
functions.
We show that, due to the relation~\mref{w7}, $W(x)$\ are again expressed
in terms of Calogero polynomials. Therefore, we cannot in this way gain
much
insight into their structure. Second, we extend an old result (see, for 
example, Ref.\cite{b15}) concerning the $sL(2,\rr)$\ structure of 
Calogero--Moser model, i.e. Calogero model without harmonic term. It
appears
that this structure can be easily describe using the representation
theory
in the basis diagonalizing noncompact generators~\cite{b17}.
\section{Laguerre polynomials}
Laguerre polynomial $L_n^\alpha$\ is defined to be the polynomial
solution
to the equation~\cite{b16}
\be
z{d^2u\over dz^2}+(\alpha-z+1){du\over dz}+nu&=&0\label{w10}
\ee
normalized according to the condition
\be
u(z)&=&(-1)^n{z^n\over n!}+\mbox{\ \ lower degree terms}\label{w11}
\ee

Let us rewrite \mref{w10} in form
\be
\left[z{d\over dz}-\left((\alpha+1){d\over dz}+z{d^2\over
dz^2}\r)\r]u&=&nu\label{w12}
\ee
Put
\be
A&\equiv&(\alpha+1){d\over dz}+z{d^2\over dz^2}\label{w13}
\ee
and
\be
u&=&e^{-A}v\label{w14}
\ee

Simple calculation gives equation for $v$,
\be
z{dv\over dz}&=&nv,\label{w15}
\ee
so that we get the following representation for Laguerre polynomials
\be
L_n^\alpha(z)&=&{(-1)^n\over n!}e^{-(\alpha+1){d\over dz}-z{d^2\over
dz^2}}z^n\label{w16}
\ee
\section{The Calogero wave function}
In order to find the representation~\mref{w5} for Calogero wave
functions 
we have to calculate
\be
W(x)&=&e^{{1\over 4\omega}O_L}\phi_{nk}\label{w17}
\ee
Following Calogero~\cite{b1} we introduce spherical coordinates in the
space
of relative coordinates. Since $\phi_{nk}$\ depends only on relative 
coordinates one can neglect the center--of--mass coordinate dependence
of $O_L$; then $O_L$\ can be rewritten as
\be
O_L&=&r^{2-N}{\p\over \p r}\left(r^{N-2}{\p\over\p r}\r)+aN(N-1)
{1\over r}{\p\over\p r}+{1\over r^2}(\hat{L}+2a\hat{M});\label{w18}
\ee
$\hat{L}+2a\hat{M}$\ is an angular part and the following equation holds
\be
(\hat{L}+2a\hat{M})(r^{-k}P_k(x))&=&-k(k+N-3+aN(N-1))r^{-k}P_k(x)\label{w19}
\ee
Therefore
\be
W(x)&=&\exp\left({1\over 4\omega}\left(r^{2-N}{\p\over\p
r}(r^{N-2}{\p\over\p r})
+aN(N-1){1\over r}{\p\over\p r}\r)+{1\over 4\omega r^2}\r.\nn\\
&&\left.(\hat{L}+2a\hat{M})\r)
\left(L^b_n(\omega r^2)r^k\r)\left(r^{-k}P_k(x)\r)=\nn\\
&&=r^{-k}P_k(x)\exp\left({1\over 4\omega}\left(r^{2-N}{\p\over\p
r}(r^{N-2}
{\p\over\p r})+aN(N-1){1\over r}{\p\over\p r}\r.\r.\nn\\
&&\left.\left.-\frac{k(k+N-3+aN(N-1))}{r^2}\r)\r)\left(L^b_n(\omega
r^2)r^k\r)
\label{w20}
\ee

Put $z=\omega r^2$; the relevant part of the $rhs$\ of eq.\mref{w20}
reads
\be
&&z^{-{k\over 2}}\exp\left\{z{\p^2\over\p z^2}+\left({N-1+aN(N-1)\over
2}\r)
{\p\over \p z}-\r.\nn\\
&&-\left.{k(k+N-3+aN(N-1)\over 4z}\r\}z^{k\over 2}=\nn\\
&&=\exp\left\{z^{-{k\over 2}}\left(z{\p^2\over \p
z^2}+\left({N-1+aN(N-1)\over
2}\r){\p\over\p z}-\r.\r.\nn\\
&&-\left.\left.{k(k+N-3+aN(N-1)\over 4z}\r)z^{k\over 2}\r\}=\nn\\
&&=\exp\left\{z{\p\over\p z^2}+\left({2k+N-1+aN(N-1)\over
2}\r){\p\over\p z}\r\}
\label{w21}
\ee

Eqs.\mref{w8},\mref{w16},\mref{w20} and \mref{w21} imply that, up to a
normalization constant,
\be
W(x)&=&r^{2n}P_k(x)\label{w22}
\ee
and
\be
\phi_{nk}&=&e^{-{1\over 4\omega}O_L}(r^{2n}P_k(x))\label{w23}
\ee

Eq.\mref{w23} provides the representation for the wave functions in
Calogero
basis.

Due to the fact that $P_k(x)$\ appears also on the right hand side of 
eq.\mref{w23} we do not gain much insight into the structure of Calogero
polynomials. This is due to the fact that the operators classifying the
polynomials $P_k(x)$\ commute with $sL(2,\rr$) generators while $O_L$\ 
can be expressed in terms of them.

\section{The $sL(2,\rr)$\ structure of Calogero--Moser eigenfunctions}
Let us consider the Calogero--Moser model obtained from eq.\mref{w1}
by putting $\omega=0$.
\be
H_{CM}&=&\sum\limits_i p^2_i/2+{a(a-1)\over
2}\sum\limits_{i\not=j}{1\over
(x_i-x_j)^2}\label{w24}
\ee

Obviously, $H_{CM}$\ has only continuous spectrum (scattering states).
The relevant wave function read~\cite{b1}
\be
\Psi_{p_k}&=&\left(\prod\limits_{i<j}(x_i-x_j)\r)^a
r^{-b}J_b(pr)P_k(x)\label{w25}
\ee
and correspond to the energies $E=p^2/2$.

The solutions~\mref{w25} can be classified according to the
representations
of $sL(2,\rr$) algebra.

Define
\be
J_+&=&\fr\left(\sum\limits_i p_i^2/2+{a(a-1)\over 2}\sum\limits_{i\not=
j}
{1\over(x_i-x_j)^2}\r)\equiv\fr H_{CM}\nn\\
J_-&=&-\sum\limits_i x_i^2\label{w26}\\
J_2&=&{1\over 4}\sum\limits_i(x_ip_i+p_ix_i);\nn
\ee
then
\be
{}[J_2,J_\pm]&=&\pm iJ_\pm\label{w27}\\
{}[J_+,J_-]&=&2iJ_2\nn
\ee

Let us gauge out the factor ($\prod\limits_{i<j}(x_i-x_j))^a$\ and pass
to the 
center--of--mass ($R\equiv{1\over N}\sum\limits_ix_i$) and relative 
coordinates~\cite{b1}. We get
\be
J_+&=&-{1\over 4N}{\p^2\over\p R^2}-{1\over 4}\left[ r^{2-N}{\p\over \p
r}
\left(r^{N-2}{\p\over\p r}\r)+{N(N-1)a\over r}{\p\over\p r}+\r.\nn\\
&&\left.+{1\over r^2}(\hat{L}+2a\hat{M})\r]\nn\\
J_-&=&-(r^2+NR^2)\label{w28}\\
J_2&=&\left(-{i\over 2}R{\p\over\p R}-{i\over 4}\r)+\left(-{i\over 2}r
{\p\over\p r}-i({N-1\over 4})-{iaN(N-1)\over 4}\r);\nn
\ee

Actually, our algebra is rather diagonal part of $sL(2,\rr)\oplus
sL(2,\rr)$.
However, the center--of--mass part may be ignored and we are left with
$sL(2,\rr)$\ algebra in relative coordinates.

Let us now recall the structure of $D^{(+)}$\ represenations of
$sL(2,\rr)$\ 
in the basis diagonalizing $J_2$~\cite{b17}. The action of generators
reads
\be
J_2|\lambda>&=&\lambda|\lambda>\nn\\
J_+|\lambda>&=&h(\lambda+i)|\lambda+i>\label{w29a}\\
J_-|\lambda>&=&-|\lambda-i>\nn
\ee
where
\be
h(\lambda)&=&\lambda(\lambda-i)-j(j+1)\label{w29b}
\ee

Contrary to the continuous nonexceptional series for discrete
representation
$D^{(+)}$\ the multiplicity of any $\lambda$\ is one~\cite{b17}. It is
not
difficult to see that $J_2$\ is dilatation--type operator and the
$D^{(+)}$\ 
representation can be related to harmonic analysis on $\rr_+$. In fact,
consider the Hilbert space of functions on $\rr_+$\ equipped with a
scalar
product
\be
(f,g)&=&\int\limits^\infty_0 dr r^c
\stackrel{\mbox{---}}{f(r)}g(r)\label{w30}
\ee

Define the action of unitary dilatation group as
\be
\left(U(\alpha)f\r)(r)&=&\left(e^{\alpha\over
2}\r)^{c+1}f(e^{\alpha\over 2}r)\label{w31}
\ee
Putting
\be
U(\alpha)&=&e^{i\alpha D}\label{w32}
\ee
one gets
\be
D&=&-{i\over 2}r{d\over dr}-{(c+1)i\over 4}\label{w33}
\ee

$D$\ has purely continuous spectrum covering the whole real axis. The 
conditions
\be
Df_\lambda&=&\lambda f_\lambda\label{w34}\\
(f_\lambda,f_{\lambda'})&=&\delta(\lambda-\lambda')\nn
\ee
imply
\be
f_\lambda (r)&=&{1\over\sqrt{\pi}}r^{2i\lambda-({c+1\over
2})}\label{w35}
\ee

Modulo typical mathematical subtleties (cf. the theory of Fourier
transform)
one can write any element $\Psi$\ of the Hilbert space as follows
\be
\Psi(r)&=&\int\limits^\infty_{-\infty}
d\lambda\tilde{\Psi}(\lambda)f_\lambda
\equiv{1\over\sqrt{\pi}}\int\limits^\infty_{-\infty}
d\lambda\tilde{\Psi}
(\lambda)r^{2i\lambda-({c+1\over 2})}\label{w36}
\ee
where
\be
\tilde{\Psi}(\lambda)&=&(f_\lambda,\Psi(r))={1\over\sqrt{\pi}}
\int\limits^\infty_0 drr^{c-1\over 2}r^{-2i\lambda}\Psi(r)\label{w37}
\ee
Eqs.\mref{w36},\mref{w37} represent nothing but Mellin transformation.

Now, identifying
\be
|\lambda>&\equiv&f_\lambda,\;\;\;\;D\equiv J_2\label{w38}
\ee
one can find the action of $sL(2,\rr)$\ generators. First, the action of
$J_2,J_\pm$\ on wave functions $\tilde{\Psi}(\lambda)$\ reads
\be
J_2\tilde{\Psi}(\lambda)&=&\lambda\tilde{\Psi}(\lambda)\nn\\
J_-\tilde{\Psi}(\lambda)&=&-\tilde{\Psi}(\lambda+i)\label{w39}\\
J_+\tilde{\Psi}(\lambda)&=&h(\lambda)\tilde{\Psi}(\lambda-i)\nn
\ee
Using eq.\mref{w37} one can rewrite the action of $J$'s in terms of 
r--dependent wave functions:
\be
(J_-\Psi)(r)&=&-r^2\Psi(r)\label{w40}\\
(J_+\Psi)(r)&=&\left(-{1\over 4}{d^2\over dr^2}-{c\over 4}{1\over r}
{d\over dr}-{j(j+1)\over r^2}-{(c+1)(c-3)\over 16r^2}\r)\Psi(r)\nn
\ee

In our case
\be
c&=&(N-2)+aN(N-1);\label{w41}
\ee
the first term on the right hand side comes from the Jacobian of 
transformation to spherical coordinates in the space of relative
coordinates
while the second one is the result of gauging out the
$(\prod\limits_{i<j}
(x_i-x_j))^a$\ factor.

In order to compare eqs.\mref{w28} and \mref{w40} we insert
eq.\mref{w19}
into \mref{w28}. The result coincides with eq.\mref{w40} provided
\be
j&=&{k\over 2}+{N-5+aN(N-1)\over 4}\label{w42}
\ee

We conclude that, given the Calogero polynomial $P_k(x)$, the wave 
functions~\mref{w35} span the $D^{(+)}$\ representation of our
$sL(2,\rr)$\ 
algebra. It is easy to show that diagonalizing the $J_3(=\fr(J_+-J_-))$\ 
generator we reveal the $sL(2,\rr)$\ structure of the wave functions
of Calogero model~\cite{b15}.

\end{document}